\documentclass[apl,showpacs,showkeys,amsmath,amssymb,twocolumn]{revtex4-1}
\usepackage[portuges]{babel} 
\usepackage[applemac]{inputenc}
\usepackage[T1]{fontenc} 
\usepackage{makeidx} 
\usepackage{graphicx} 
\usepackage{dcolumn} 
\usepackage{array} 
\usepackage{amssymb} 
\usepackage{amsmath}
\usepackage{textcomp}
\usepackage{rotating}
\usepackage{wasysym}
\usepackage{multirow}
\usepackage{subfigure}
\usepackage{eucal}
\usepackage{mathrsfs}
\usepackage{units}
\usepackage{rotating}
\usepackage[all]{xy}
\usepackage{float} 
\usepackage{amsmath} 
\usepackage{amsfonts} 
\usepackage{bm} 
\usepackage[amssymb]{SIunits}

\begin{document}

\title{Oscillating magnetocaloric effect of a 2D non-relativistic diamagnetic material}

\author{M.S. Reis}\email{marior@if.uff.br}\affiliation{Instituto de F\'{i}sica, Universidade Federal Fluminense, Av. Gal. Milton Tavares de Souza s/n, 24210-346, Niter\'{o}i-RJ, Brasil}

\date{\today}

\begin{abstract}
Among the magnetic materials, those with ferromagnetic character are, by far, the most studied in what concerns applications of the magnetocaloric effect. However, recently, diamagnetic materials received due attention never received before, and an oscillatory behavior, analogous to the de Haas-van Alphen effect, has been found. The present effort describes in details the magnetocaloric properties of a 2D non-relativistic material (a Gold thin film, for instance), where oscillations, depending on the reciprocal magnetic field $1/B$, are found. A comparison of the magnetic entropy change per electron for some cases is presented and we found $\approx10^{-1}$ k$_{\text{B}}$ (@109.3 K) for graphenes, $\approx10^{-5}$ k$_{\text{B}}$ (@0.7 K) for 2D Gold and $\approx 10^{-7}$ k$_{\text{B}}$ (@0.7 K) for 3D Gold.
\end{abstract}

\maketitle
\section{Introduction}

The magnetocaloric effect is an interesting property in which magnetic materials, under a magnetic field change, are able to exchange heat with a thermal reservoir (considering an isothermal process), or even change its temperature (considering an adiabatic process). This effect is completely analogous to the compression-expasion thermal-mechanical cycle; and therefore the main purpose of the scientific community is to built a thermo-magnetic machine, to substitute, in a near future, those standard, non-economical and non-environmental friendly Freon-like refrigerators\cite{tishin_book}.

Thus, the scientific community dealing with caloric effects (baro-, magneto- and electro-), focused attention to applied purposes and therefore, to this end, ferro-materials are the most useful, since near phase transitions the caloric effects are maximized\cite{tishin_book}. However, recently, the magnetic entropy change $\Delta S(T,\Delta B)$ and the adiabatic temperature change $\Delta T(T,\Delta B)$ of 3D non-relativistic diamagnetic materials have been described in details\cite{MCE_OSC_DS,MCE_OSC_DT}. An oscillatory behavior was found, due to the crossing of the Landau levels with the Fermi level, analogously to the de Haas-van Alphen effect\cite{greiner}. This new (oscillating) magnetocaloric effect has interesting features and, as suggested, can be used as sensible magnetic field sensor\cite{MCE_OSC_DS}.

However, in spite to be interesting, this oscillating effect occurs at high values of magnetic field (few or more Teslas) and low temperatures (c.a. 1 K). In order to further understand and optimize the effect, a relativistic 2D material (a graphene) was studied\cite{mce_grafeno}; and, surprisingly, a very similar oscillations were found, however, at very comfortable values of temperature: c.a. 100 K (and few Teslas). To deeper understand the magnetocaloric properties of graphenes, more recently the influence of a longitudinal electric field was described\cite{graphene_MCE_electric} and verified that the caloric potentials can be easily rules by this electric field. The electrocaloric effect of graphens was also explored\cite{ece_grafeno}.

What does happen for a material that has low dimension (2D), like graphene, but is non-relativisit in nature, as the first case above described? This is the aim of the present effort, to present the magnetic entropy change of a 2D non-relativistic film. To this purpose, we derive the density of states (details are in the appendix), and then, from known thermodynamic relationships, the magnetic entropy change is obtained.

\section{Magnetic entropy change}

The magnetic entropy $S(T,B)$, as a function of temperature $T$ and magnetic field $B$, depends on the \emph{grand} canonical partition function $\mathcal{Z}(T,B)$ through:
\begin{equation}\label{entropia}
S(T,B)=k_B\frac{\partial}{\partial T}\left[T\ln\mathcal{Z}(T,B)\right]
\end{equation}
where
\begin{equation}\label{zini}
\ln\mathcal{Z}(T,B)=\int_0^\infty d\varepsilon\; g_1(\varepsilon)\;\ln\left(1+ze^{-\beta\varepsilon}\right),
\end{equation}
$z=e^{\mu\beta}$ is the fugacity, $\mu$ is the chemical potential and $\beta=1/k_BT$. In addition, $g_1(\varepsilon)$ is the one-particle density of states, given by (see Appendix):
\begin{equation}\label{g1}
g_1(\varepsilon)=g_0+g_B(\varepsilon)
\end{equation}
where
\begin{equation}
g_0=\frac{L^2}{4\pi}\frac{2m}{\hbar^2}
\end{equation}
is the zero-field density of state (with no spin degeneracy),  and
\begin{equation}\label{gbgb}
g_B(\varepsilon)=2g_0\sum^{+\infty}_{l=1}(-1)^l\cos\left(\frac{l\pi}{\mu_BB}\varepsilon\right)
\end{equation}

From equation \ref{g1}, it is easy to see that the logarithm of the \emph{grand} partition function has two contributions:
\begin{equation}\label{zomzb}
\ln\mathcal{Z}(T,B)=\ln\mathcal{Z}_0(T)+\ln\mathcal{Z}_B(T)
\end{equation}
The first one gives the zero-field entropy and, as will be shown further in this text, does not contribute to the magnetic entropy change. Thus, let us focus our attention to the second term that, from equation \ref{gbgb}, is:
\begin{equation}\label{rerere}
\ln\mathcal{Z}_B(T)=\int_0^\infty d\varepsilon\, 2g_0\sum^{+\infty}_{l=1}(-1)^l\cos\left(\frac{l\pi}{\mu_BB}\varepsilon\right)
\ln\left(1+ze^{-\beta\varepsilon}\right)
\end{equation}

After integration by parts (two times), we achieve:
\begin{align}\label{lnzb}\nonumber
\ln\mathcal{Z}_B(T)=2g_0\sum^{+\infty}_{l=1}(-1)^l&\left\{G(\varepsilon)|^\infty_0+\left(\frac{\mu_BB\beta}{l\pi}\right)\mathcal{G(\varepsilon)}|_0^\infty\right.\\
&\left.-\frac{1}{4}\left(\frac{\mu_BB\beta}{l\pi}\right)^2\mathcal{I}(T,B)\right\}
\end{align}
where
\begin{equation}
G(\varepsilon)=\frac{\mu_BB}{l\pi}\sin\left(\frac{l\pi}{\mu_BB}\varepsilon\right)\ln\left(1+ze^{-\beta\varepsilon}\right),
\end{equation}
\begin{equation}
\mathcal{G}(\varepsilon)=-\frac{\mu_BB}{l\pi}\cos\left(\frac{l\pi}{\mu_BB}\varepsilon\right)\frac{1}{z^{-1}e^{\beta\varepsilon}+1}
\end{equation}
and
\begin{align}\nonumber
\mathcal{I}(T,B)=&\int_0^\infty\frac{\cos\left(\frac{l\pi}{\mu_BB}\varepsilon\right)}{\cosh^2\left(\frac{\beta(\varepsilon-\mu)}{2}\right)} d\varepsilon\\
&\xrightarrow{\varepsilon_F\gg k_BT}\frac{4\pi^2l}{\mu_BB\beta^2}\frac{\cos\left(\frac{l\pi}{\mu_BB}\varepsilon_F\right)}{\sinh\left(\frac{l\pi^2}{\mu_BB\beta}\right)}
\end{align}
Above, the condition $\varepsilon_F\gg k_BT$ implies to $\mu\sim \varepsilon_F$ and therefore $z\gg1$. As a consequence, $G(\infty)$, $G(0)$ and $\mathcal{G}(\infty)$ are zero; and
\begin{equation}
\mathcal{G}(0)\rightarrow-\frac{\mu_BB}{l\pi}
\end{equation}

Equation \ref{rerere} can then be written with two contributions:
\begin{equation}\label{twoono}
\ln\mathcal{Z}_B(T)=\ln\mathcal{Z}_B^{no}(T)+\ln\mathcal{Z}_B^o(T)
\end{equation}
where
\begin{equation}\label{nonoon}
\ln\mathcal{Z}_B^{no}(T)=-\frac{1}{6}g_0\beta(\mu_BB)^2
\end{equation}
has a non-oscillatory character and
\begin{equation}\label{ooo}
\ln\mathcal{Z}_B^o(T)=-2g_0\mu_BB\sum^{+\infty}_{l=1}\frac{(-1)^l}{l}\cos\left(\frac{l\pi}{\mu_BB}\varepsilon_F\right)\frac{1}{\sinh(x_l)}
\end{equation}
has an oscillating behavior. Above, $x_l=lx$ and
\begin{equation}
x=\pi^2\frac{k_BT}{\mu_BB}
\end{equation}

Thus, equation \ref{zomzb} reads as:
\begin{equation}
\ln\mathcal{Z}(T,B)=\ln\mathcal{Z}_0(T)+\ln\mathcal{Z}_B^{no}(T)+\ln\mathcal{Z}_B^{o}(T)
\end{equation}
where the second and third terms are given by equations \ref{nonoon} and \ref{ooo}, respectively. From equation \ref{entropia}, it is possible to obtain the entropy of the system, that reads as:
\begin{equation}
S(T,B)=S_0(T)+S_B^o(T)
\end{equation}
since $S_B^{no}=0$. However, we are interested to obtain the magnetic entropy change $\Delta S(T,\Delta B)=S(T,B)-S(T,0)$ that, on its turn, is:
\begin{equation}\label{finall}
\Delta S(T,\Delta B)=S_B^o(T)=-2Nk_B\frac{1}{n}\cos(n\pi)\mathcal{T}(x)
\end{equation}
since $S_{B=0}^o(T)=0$. Above,
\begin{equation}\label{n3n}
n=\frac{\varepsilon_F}{\mu_BB}
\end{equation}
and, for convenience, we re-write $x=\pi^2tn$, where
\begin{equation}
t=\frac{k_BT}{\varepsilon_F}
\end{equation}
In addition,
\begin{equation}\label{tau}
\mathcal{T}(x)=\frac{xL(x)}{\sinh(x)}
\end{equation}
and $L(x)$ is the Langevin function:
\begin{equation}
L(x)=\coth(x)-\frac{1}{x}
\end{equation}
As usually considered for thermodynamic quantities with this kind of oscillations\cite{greiner, MCE_OSC_DS}, we considered only $l=1$ term: the hyperbolic sine function at the denominator of equation \ref{tau} rapidly damping the summation. Of course, it is an approximation; however, after this assumption, we can understand the physics behind this model.

\begin{figure}
\begin{center}
\includegraphics[width=10cm]{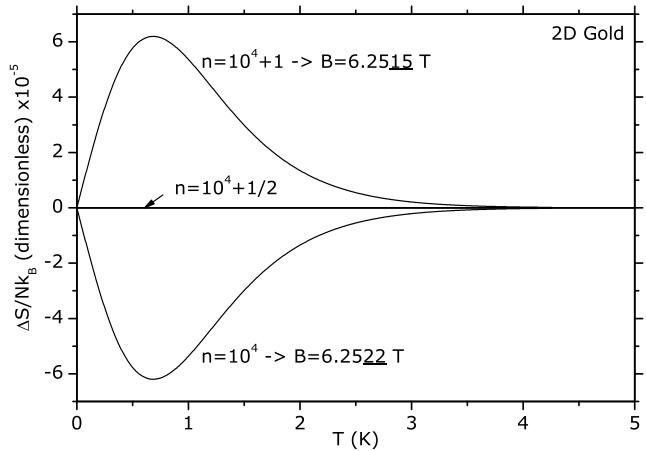}
\end{center}
\caption{Oscillating magnetic entropy change per electron as a function of temperature, for a 2D Gold film. Note only 0.7 mT of change on the magnetic field change is enough to invert the caloric effect from normal/negative to inverse/positive.} \label{fig1}
\end{figure}

Let us consider a Gold thin film, with Fermi energy $\varepsilon_F=3.62$ eV. From equation \ref{n3n} it is possible to see that $B(n=1)=6.2522\times10^4$ T and therefore completely out of a laboratory range. On the other hand, $B(n=10^4)=6.2522$ T, and therefore this is the order of magnitude of $n$ we must consider. Note the magnetic field change needed to invert the magnetic entropy change (from normal/negative to inverse/positive) is $B(n=10^4+1)=6.2515$ T, i.e., a difference of 0.7 mT. The temperature dependence of the magnetic entropy change lies on the $\mathcal{T}(x)$ function and this behavior is presented in figure \ref{fig1}. This function peaks in $x=\pi^2tn=1.6$ and therefore, considering $n=10^4$, the maximum magnetic entropy change occurs at $T=0.7$ K.

\begin{figure}
\begin{center}
\includegraphics[width=8cm]{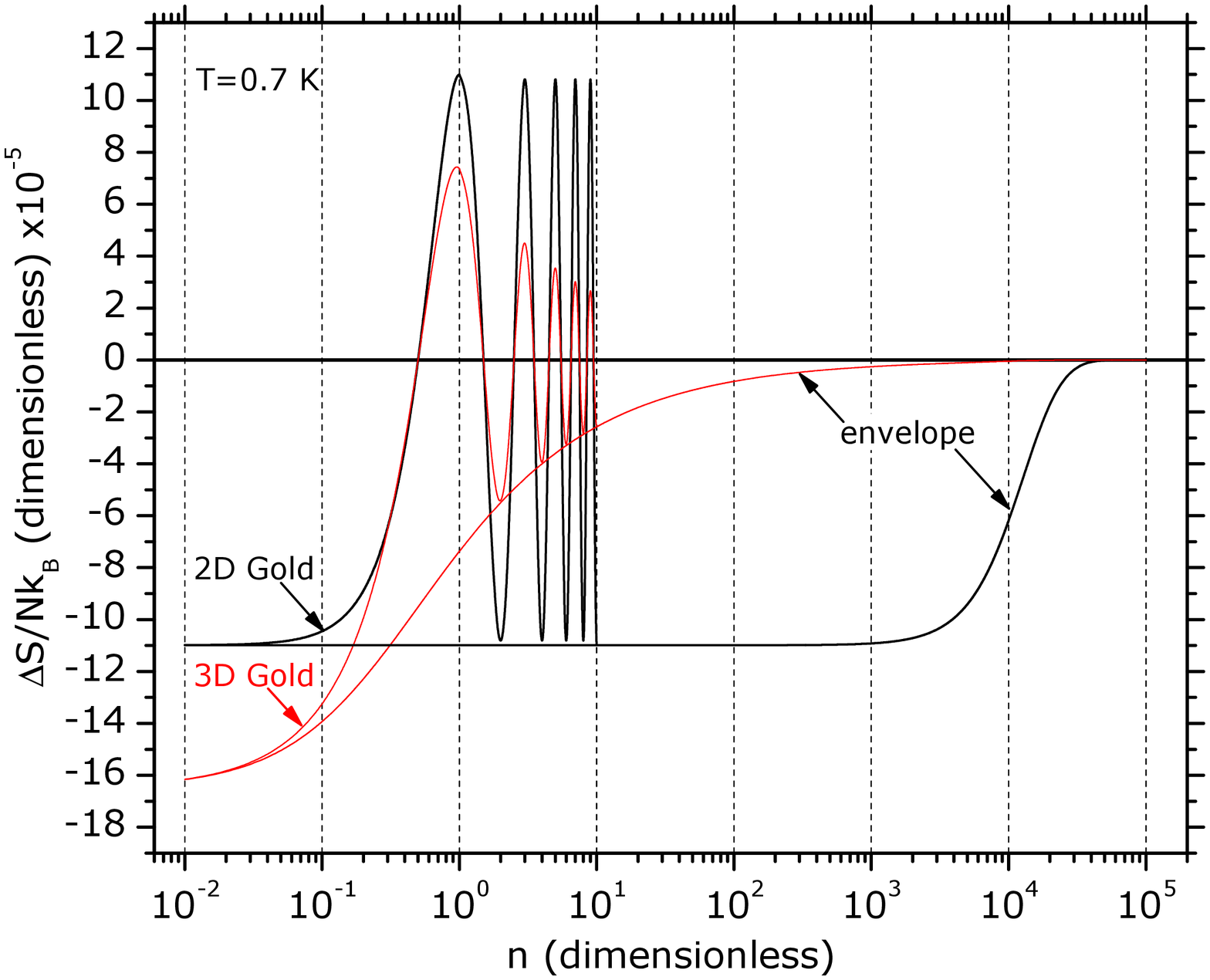}
\end{center}
\caption{Oscillating magnetic entropy change per electron as a function of $n$, inversely proportional to the magnetic field $B$. The oscillatory behavior is evident and has a remarkable difference comparing 2D and 3D models. The \emph{envelope} rules the amplitude of the oscillations. For the sake of clearness of the figure, above $n=10$ only the \emph{envelope} is shown; and the oscillations are suppressed.} \label{fig2}
\end{figure}

The dependence of the magnetic entropy change as a function of $n$ is presented in figure \ref{fig2}, for that temperature that maximizes this effect (considering $n=10^4$), i.e., 0.7 K. It has an oscillatory behavior due to the cosine term on equation \ref{finall} (note the \emph{envelope} of these oscillations is equation \ref{finall} without the cosine term).

\section{Comparison with a 3D non-relativistic material}

The magnetocaloric effect of diamagnetic materials begun to be discussed very recently, considering a 3D non-relativistic Gold specimen\cite{MCE_OSC_DS}. After a similar evaluation as these presented in this contribution, the magnetic entropy change (as well as the adiabatic temperature change\cite{MCE_OSC_DT}), has been obtained\cite{MCE_OSC_DS}:
\begin{equation}\label{nnn}
\Delta S(T,\Delta B)
=-\frac{3}{2}Nk_B\left(\frac{1}{n+1/4}\right)^{3/2}\cos(n\pi)\;\mathcal{T}(y)
\end{equation}
where
\begin{equation}\label{nmesmo}
n=\frac{\varepsilon_F}{\mu_BB}-\frac{1}{4},
\end{equation} 
\begin{equation}\label{x3d}
y=\pi^2t\left(n+\frac{1}{4}\right)
\end{equation}
and $\mathcal{T}(y)$ given by equation \ref{tau}.

Note the similarity between equation \ref{nnn} (3D case) and equation \ref{finall} (2D case). In what concerns the temperature dependence of the magnetic entropy change, both are driven by $\mathcal{T}(y)$. Since the argument of this function is (almost) the same for both cases, then $\mathcal{T}(x)$ peaks approximately at the same value of temperature (0.7 K), for the 2D and 3D models. Thus, figure \ref{fig1} is a good approximation for the behavior of the magnetic entropy change as a function of temperature for the 3D non-relativistic case; except for the fact the 2D case is almost 2 orders of magnitude bigger than the 3D case (for $n=10^4$) - this fact will be clearer further in the text.

On the other hand, the magnetic entropy change as a function of $n$ has important differences. In spite of the period of oscillations be (almost) the same for both cases (the differences lie on the Fermi energy values and the 1/4 factor-see equation \ref{nmesmo} and \ref{n3n}), the \emph{envelope} of these oscillations are remarkable different - the \emph{envelope} is the entropy change without the cosine terms. Note the maximum amplitude for the 3D case rapidly goes to zero, while for the 2D case it keeps almost constant and then, for large values of $n$, suddenly goes to zero. However, as discussed above, $n=10^4$ corresponds to $B=6.2522$ T, a range of magnetic field close to standard laboratory values; and therefore, for $n=10^4$, the magnetic entropy change, per electron, for a 2D Gold film is two orders of magnitude bigger than the 3D Gold case.

\section{Comparison with a 2D relativistic material: graphenes}

More recently, the caloric properties of graphenes have been described\cite{graphene_MCE_electric,ece_grafeno,mce_grafeno,epl_quantum}; in a similar fashion as presented above. The magnetic entropy change per graphene area was evaluated and reads as\cite{mce_grafeno}:
\begin{equation}\label{DSFIM}
\Delta S(T,\Delta B)=2 k_B\frac{N_0}{m}\cos(\pi m)\mathcal{T}(z)
\end{equation}
where $N_0=10^{16}$ m$^{-2}$ is the density of charge carriers\cite{NN,JAP},
\begin{equation}\label{mmesmo}
m=N_0\frac{\phi_0}{B},
\end{equation}
$\phi_0=\pi\hbar/e=2.06\times10^{-15}$ Tm$^2$ is the magnetic flux quantum, 
\begin{equation}
z=\frac{m}{N_0}\frac{k_BT}{\tilde{v}_F},
\end{equation}
$\tilde{v}_F=\hbar v_F/2\pi\sqrt{N_0\pi}=9.43\times10^{-38}$Jm$^2$ and $v_F=10^6$ m/s is the Fermi velocity. Note both, $m$ and $x$ are dimensionless. Interestingly, $\mathcal{T}(z)$ is also given by equation \ref{tau}.

\begin{figure}
\begin{center}
\includegraphics[width=8cm]{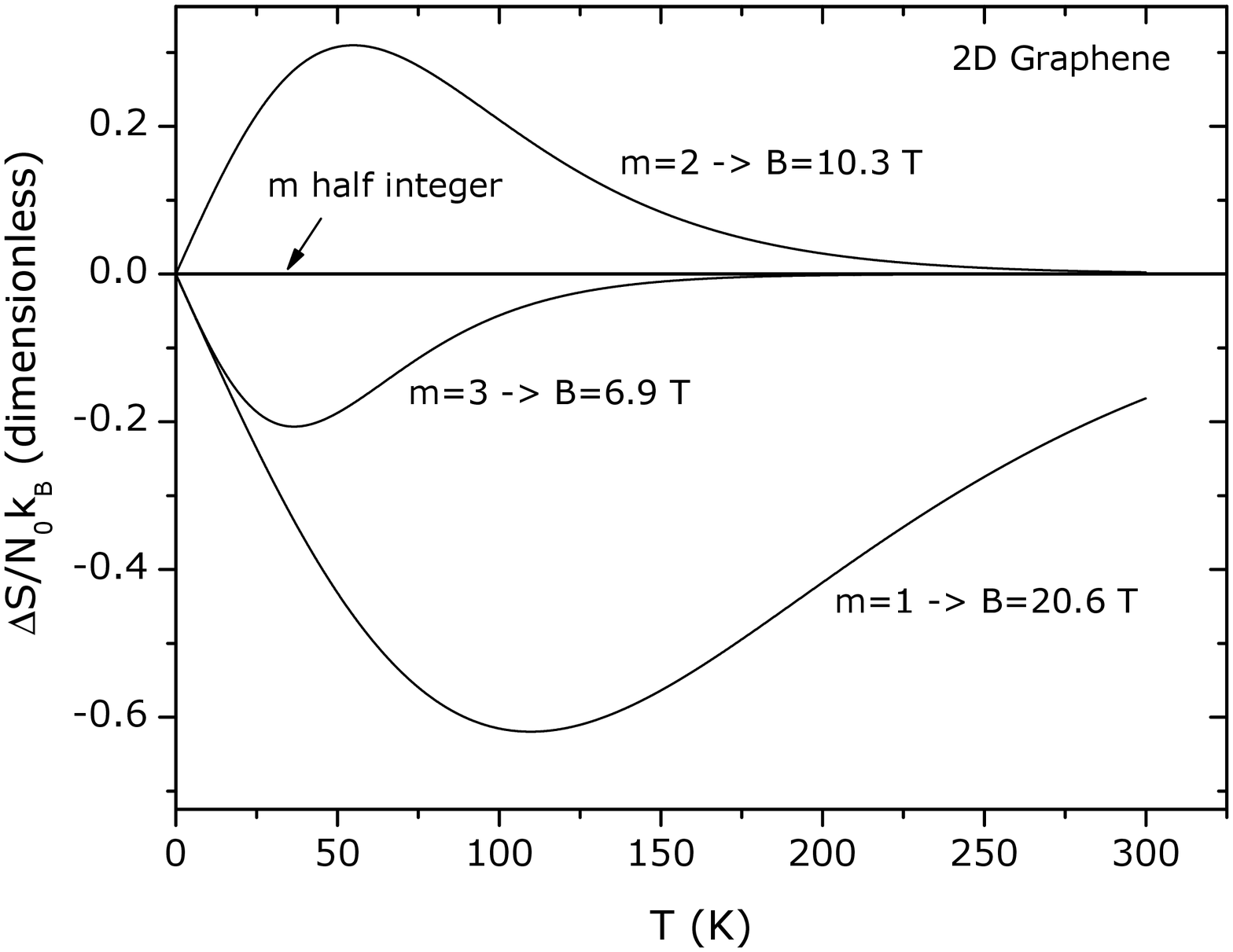}
\end{center}
\caption{Oscillating magnetic entropy change per electron for a graphene, as a function of temperature.} \label{fig3}
\end{figure}

The entropy change of graphenes (equation \ref{DSFIM}), provides a huge difference to what was discussed until this point. Note $B(m=1)=20.6$ T and $B(m=3)=6.9$ T, i.e., within the laboratory range. Thus, for this graphene comparison, we will focus to $m\approx1$, instead of those $n=10^4$ for a non-relativistic material. In addition, the temperature dependence of this entropy change for graphenes is also on  $\mathcal{T}(z)$ function of equation \ref{tau}, that peaks at $z=1.6$ and therefore, for $m=1$, the maximum magnetic entropy change occurs at T=109.3 K. Figure \ref{fig3} summarizes these considerations and, comparing with figure \ref{fig1}, it is obvious the huge difference, mainly in the temperature in which the oscillating behavior occurs.

\begin{figure}
\begin{center}
\includegraphics[width=8cm]{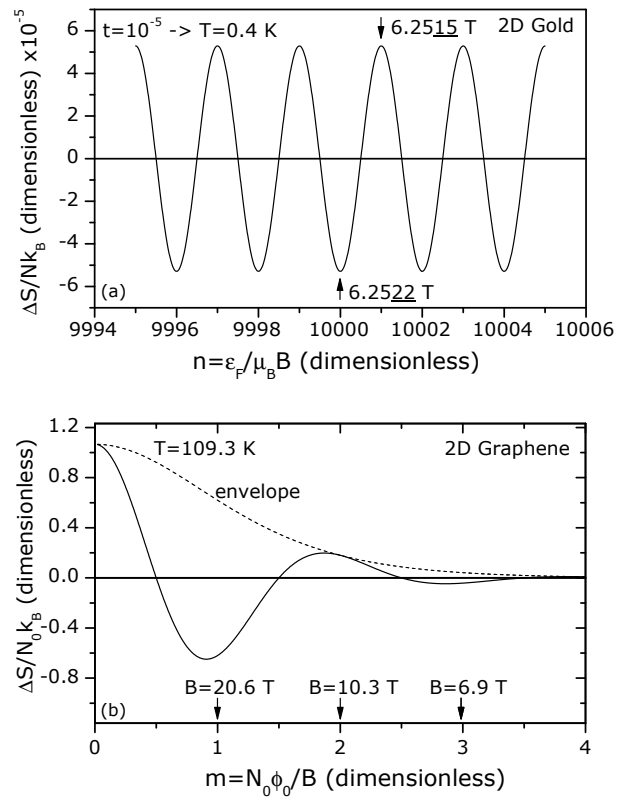}
\end{center}
\caption{Oscillating magnetic entropy change per electron for a (top) 2D non-relativistic film-Gold and (bottom) graphene, as a function of the inverse magnetic field.} \label{fig4}
\end{figure}

Analogously to figure \ref{fig2}, the entropy change as a function of $m$ for graphenes is presented in figure \ref{fig4}-bottom, for values of magnetic field within the laboratory range. Note the oscillations rapidly goes to zero and few periods of oscillations can be observed. The most interesting point is the temperature and magnetic field in which the effect is; i.e., completely reasonable in what concerns a standard laboratory. For the sake of comparison, the entropy change as a function of $n$ (see equation \ref{finall}) is presented in figure \ref{fig4}-top, for a 2D non-relativisitc material. 

These huge differences came from the relativistic energy spectra for the graphene, in which the Landau levels goes proportionally to $\sqrt{B\,j}$, where $j$ is the Landau level index; while, as known and shown in Appendix, for a non-relativistic case, the energy spectra mimics an harmonic oscillator.

As a final remark, at standard laboratory values of magnetic field change (c.a. 6 T), the entropy change per electron for a 2D relativistic material (graphene), is of the order of $10^{-1}$ k$_{\text{B}}$ (@109.3 K), while for a 2D and 3D non-relativistic material (Gold, for instance), it is, respectively, of the order of $10^{-5}$ k$_{\text{B}}$ and $10^{-7}$ k$_{\text{B}}$ (@0.7 K).

\section{Conclusions}

In what concerns caloric effects, the scientific community has been focusing attention to ferro- materials, since these effects are maximized close to phase transitions. However, recently, the magnetocaloric effect of a 3D non-relativistic diamagnetic materials has been described and, surprisingly, interesting and new features arose, as, for instance, an oscillatory behavior as a function of the reciprocal magnetic field $1/B$, analogously to the de Haas-van Alphen effect. This effect was further developed considering a relativistic material: a 2D carbon sheet, i.e., a graphene; and a similar behavior was found, however, for more comfortable values of temperature (c.a. 109 K, in contrast to the low values -c.a. 1 K- found for the non-relativistic case).

In a sequence of the results mentioned above, the present effort dealt with a non-relativistic 2D material, namely a Gold thin film. We found the characteristic oscillatory behavior already mentioned for those other diamagnetic materials (compare equations \ref{finall}, \ref{nnn} and \ref{DSFIM}), however, with important aspects that must be emphasized: (i) the temperature dependence for those three different cases lies on the $\mathcal{T}(x)$ function (see equation \ref{tau}) - independently on the dimension and nature, i.e., either relativistic or non-relativistic, (ii) all cases have dependence on a cosine function, with a period proportional to the reciprocal magnetic field $1/B$ and (iii) for those 2D cases the entropy change is proportional to $B\cos(1/B)\mathcal{T}(1/B)$, while for the 3D case it depends on $B^{3/2}\cos(1/B)\mathcal{T}(1/B)$. From this, it is possible to inspect that the entropy change for a $d$-dimensional system is  $B^{d/2}\cos(1/B)\mathcal{T}(1/B)$. This proposal needs further analysis and will be published elsewhere.

\appendix

\section{Density of States}

The one-particle density of states $g_1(\varepsilon)$ follows easier from the Laplace transformation of the canonical partition function $Z_1^B(\beta)$ in the Boltzmann limit:
\begin{equation}\label{um}
g_1(\varepsilon)=\frac{1}{2\pi i}\int_{\beta-i\infty}^{\beta+i\infty}e^{\beta^\prime\varepsilon}Z_1^B(\beta^\prime)\,d\beta^\prime
\end{equation}
where, for non-interacting systems, a simple \emph{grand} canonical-canonical partition functions relationship is:
\begin{equation}\label{compara}
\ln\mathcal{Z}(T,B)=zZ_1^B(\beta).
\end{equation}
Above, $z=e^{\mu\beta}$ is the fugacity and $\mu$ the chemical potential. Thus, at this point, we need to obtain the \emph{grand} partition function in the Boltzmann limit:
\begin{equation}\label{cano}
\ln\mathcal{Z}(T,B)=\sum_nze^{-\beta\varepsilon_n}
\end{equation}
but before, the energy spectra $\varepsilon_n$ for this model. 

The Hamiltonian of this model, i.e., a 2D non-relativistic electron gas with a transversal magnetic field $\vec{B}=B\hat{k}$, can be written as:
\begin{equation}
\mathcal{H}=\frac{1}{2m}\left[p_x^2+(p_y+eBx)^2\right]
\end{equation}
and then the Schr\"{o}dinger equation reads as:
\begin{equation}\label{ho}
\left[-\frac{\hbar^2}{2m}\frac{d^2}{dx^2}+\frac{1}{2}m\omega^2(x+x_0)^2\right]\phi_n(x)=\varepsilon_n\phi_n(x)
\end{equation}
where
\begin{equation}
x_0=\frac{\hbar k_y}{eB},\;\;\;\;\;\omega=\frac{eB}{m}\;\;\;\;\textnormal{and}\;\;\;\;\;\varepsilon_n=\hbar\omega\left(n+\frac{1}{2}\right).
\end{equation}
Above, $k_y=2\pi n_y/L$ (where $n_y=0,1,2,\cdots$), is related to the translational symmetry along the $y$ axis; and the energy spectra represents the Landau levels, where $n$ is the Landau level index. Note the harmonic oscillator of equation \ref{ho} has several centers $x_0$, that depends on $n_y$ and $B$. Since these centers must be within the considered plan, i.e., $0\leq x_0\leq L$ then $0\leq n_y \leq \tilde{g}$, where $\tilde{g}=L^2eB/h$ is the degeneracy of the Landau level $n$ (note for each $n$ there are $n_y$ possibilities).

Thus, equation \ref{cano} can be rewritten as:
\begin{align}\nonumber\label{lnzzz}
\ln\mathcal{Z}(T,B)&=\sum_{n=0}^\infty z\tilde{g}\exp\left[-\beta\hbar\omega\left(n+\frac{1}{2}\right)\right]\\
&=z\frac{L^2eB}{2h}\frac{1}{\sinh(y)}
\end{align}
where $y=\mu_BB\beta$.
Thus, a simple comparison of equation \ref{lnzzz} and \ref{compara} leads to:
\begin{equation}
Z_1^B(\beta)=\frac{L^2eB}{2h}\frac{1}{\sinh(y)}
\end{equation}

From the above and equation \ref{um}, it is possible to write the density of states we are looking for:
\begin{equation}
g_1(\varepsilon)=\frac{L^2eB}{2h}\left\{\frac{1}{2\pi i}\int^{\beta+i\infty}_{\beta-i\infty}\frac{e^{\beta^\prime\varepsilon}}{\sinh(\mu_BB\beta^\prime)}d\beta^\prime\right\}
\end{equation}
that reads as:
\begin{equation}
g_1(\varepsilon)=g_0+g_B(\varepsilon)
\end{equation}
where
\begin{equation}
g_0=\frac{L^2}{4\pi}\frac{2m}{\hbar^2}
\end{equation}
is the density of state (with no spin degeneracy), of the non-perturbed 2D electron gas; and
\begin{equation}
g_B(\varepsilon)=2g_0\sum_{l=1}^{+\infty}(-1)^l\cos\left(\frac{l\pi}{\mu_BB}\varepsilon\right)
\end{equation}

\acknowledgements{This work is supported by Brazilian agencies: FAPERJ, CNPq, CAPES and PROPPi-UFF.}

\end{document}